# The study of tuning system for BEPC II*

Submitted to 'Chinese Physics C'

MI Zhenghui (米正辉)[1;2;1)] SUN Yi(孙毅)[2] PAN Weimin(潘卫民)[2] WANG Guangwei(王光伟)[2]
Li Zhongquan(李中泉)[2] DAI Jianping(戴建枰)[2] MA Qiang(马强)[2] LIN Haiying(林海英)[2]
XU bo(徐波)[2] HUANG Hong(黄泓)[2] WANG Qunyao(王群要)[2] XU Yufen(许玉芬)[2]
ZHAO Guangyuan(赵光远)[2] HUANG Tongming(黄彤明)[2] SHA Peng(沙鹏)[2]
ZHANG Xinying(张新颖)[2] MENG Fanbo(孟繁博)[1;2] LI Han(李菡)[1;2] CHEN Xu(陈旭)[1;2]
ZHAO Danyang (赵丹阳)[1;2] ZHANG Juan(张娟)[1;2] PENG Yinghua(彭应华)[1;2]

1 (University of the Chinese Academy of Sciences, Beijing 100049, China)
2 (Institute of High Energy Physics, CAS, Beijing 100049, China)

**Abstract**: Tuning system plays a very important role when superconducting cavity works. It cooperates with other control loops, to adjust the frequency of cavity with high precision, reduce the reflection power, guarantee the stability of beam and ensure the safety of superconducting cavity. This paper focuses mainly on the tuning system working principle, the working state and problems that BEPC (Beijing Electron Positron Collider) II has encountered during operation.

## 1. Introduction

Tuning system is an essential part of the SRF (Superconducting Radio Frequency) system for BEPC II, which consists of tuner and tuning loop. The frequency of BEPC II superconducting cavity is 499.8 MHz, with the temperature of 4.2 K. It works at CW (continue waves) mode, while the maximum current is 911 mA. Because of beam effects, helium pressure and microphonics, the actual frequency would deviate from the working frequency, which will result in the increase of incident and reflection power , and even cause beam loss. To ensure that the cavity works well, the frequency of cavity should be adjusted constantly by tuning system. [1][2]

BEPC II had adopted the KEKB-type tuner. By stretching the cavity to change its frequency, ensured that the cavity is always working on resonance. So the beam can run normally and also saving energy. Some key problems of BEPC II tuning system are analyzed in this article.

## 2. Mechanical structure and working principle of tuner

### 2.1 The mechanical structure of tuner

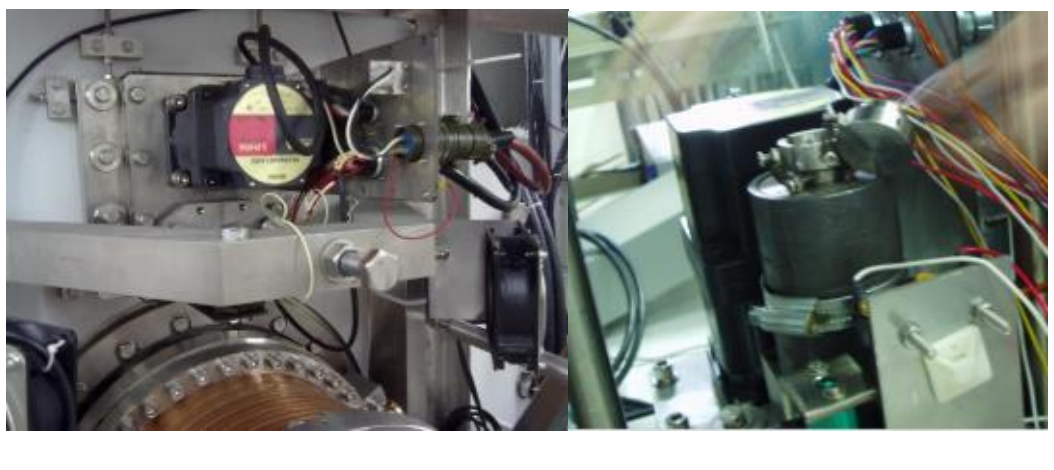

*Supported by the "500MHz superconducting cavity electromechanical tuning system", under Grant No. Y190KFEOHD

1) E-mail: mizh@ihep.ac.cn

(a) (b)

Figure1: (a) The 500 MHz tuner on cryomodule;

(b) The components of tuner

As Figure 1 (a) and (b) show, the tuner is composed of a main motor and a piezoelectric oscillator. The main motor has a large stroke in the resonant frequency, while the operation is slow. On the other hand, the piezoelectric oscillator has the small stroke and the fast operation. The piezoelectric oscillator is wrapped with a layer of 5 mm lead sheath to shield radiation. The main parameters of the tuner are listed in Table 1.[3]

Table 1: Main parameters of the tuner

| Type | Motor | Piezo |
|---|---|---|
| Tuning rate | low | fast |
| Stroke | 4mm | ~40um |
| Cavity tuning sensitivity | 330kHz/mm | |
| Tuning range | 660kHz | 6.6kHz |
| Resolution | um | several tens of nm |
| Operating temperature | room temperature | room temperature |
| Number | 1 | 1 |
| Harmonic drive ratio | 1:50 | -- |
| Type or max load | PK596AE1-H50 （0.72° /step) | -- |

Figure 2 (a) is the schematic diagram of the tuner. The motor tuner consists of a stepping motor and a reduction gear box. The reduction ratio of reduction gears is 48:72. The screw pitch of the piezoelectric oscillator support platform is 4 mm.

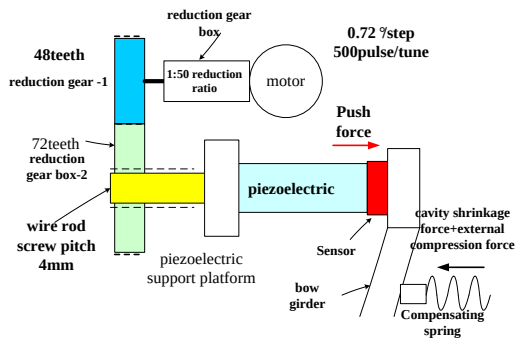

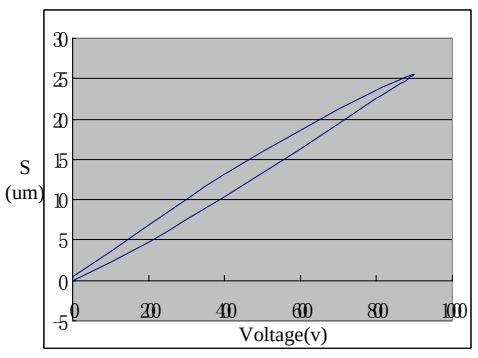

(a) (b)

Figure2 :( a) The schematic diagram of the tuner; (b) Piezoelectric oscillator voltage vs. displacement curve

The theory formula of piezoelectric oscillator support platform displacement is:

$$S=cp\times\left(\frac{0.72^{0}}{360^{0}}\right)\times\frac{1}{50}\times\frac{48}{72}\times 4 \qquad (1)$$

S: displacement of piezoelectric oscillator support platform; cp: pulse numbers

According to Formula (1), the theoretical resolution of main motor we can get is 0.1um. Through adjusting the motor driver, the displacement resolution of the motor can be improved. However, the resolution is lower than the theoretical resolution of motor tuner due to manufacturing and installation errors. Also, backlash of gears is inevitable.

Figure 2 (b) is the displacement curve of the piezoelectric oscillator during the voltage change from 0 to 900 V. The sensitivity is about 27 nm/V, hysteresis phenomenon exists during round trip.

## 2.2 The working principle of mechanical tuner

Fixture of the tuner on cryomodule is shown as Figure 3(a) and (b). The tuner arms are connected with the cavity end flange. The frequency of cavity must be lower than the working frequency in the initial state. Through preloading the cavity is adjusted to the working frequency. The preload of BEPC II cavity is about 120 kg. When the tuner works, the cavity keeps stretching. The motor and piezoelectric oscillator pushes the tuner arms to stretch cavity. According to the lever principle, the displacement of cavity is half the displacement of the tuner.

According to the cavity electromagnetic field perturbation theory, when the cavity is stretched in the axis, the frequency of the cavity will elevate. The cavity also can be simplified into a capacitor module. The frequency changes with the length of cavity.

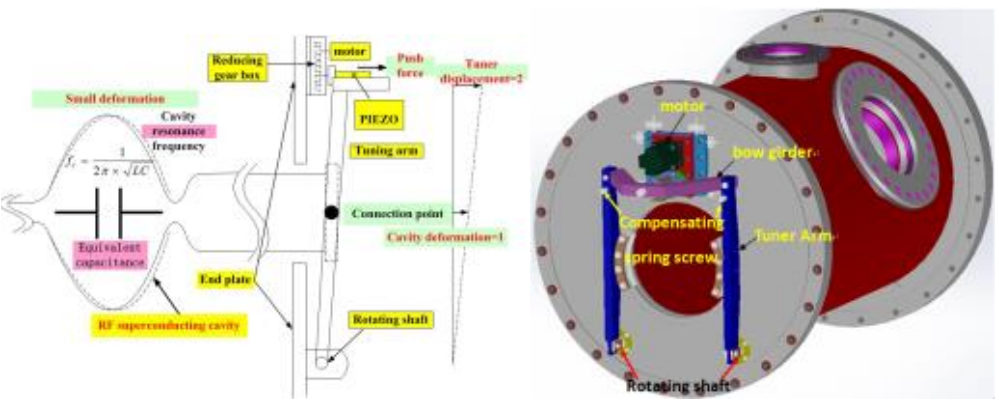

Figure3: (a) The actuator tuning diagram; (b) The 3D graph of tuner and cryomodule.

## 3. The tuning loop analysis

The tuning loop controls the frequency of the cavity based on the RF (Radio Frequency) phase error between the incident RF signal and the cavity RF field. The tuner phase error signal is processed in a servo amplifier module that provides a PI (Proportion Integration) feedback control function. The tuner controller uses the PI feedback control signal to drive the stepping motor or piezo-tuner.

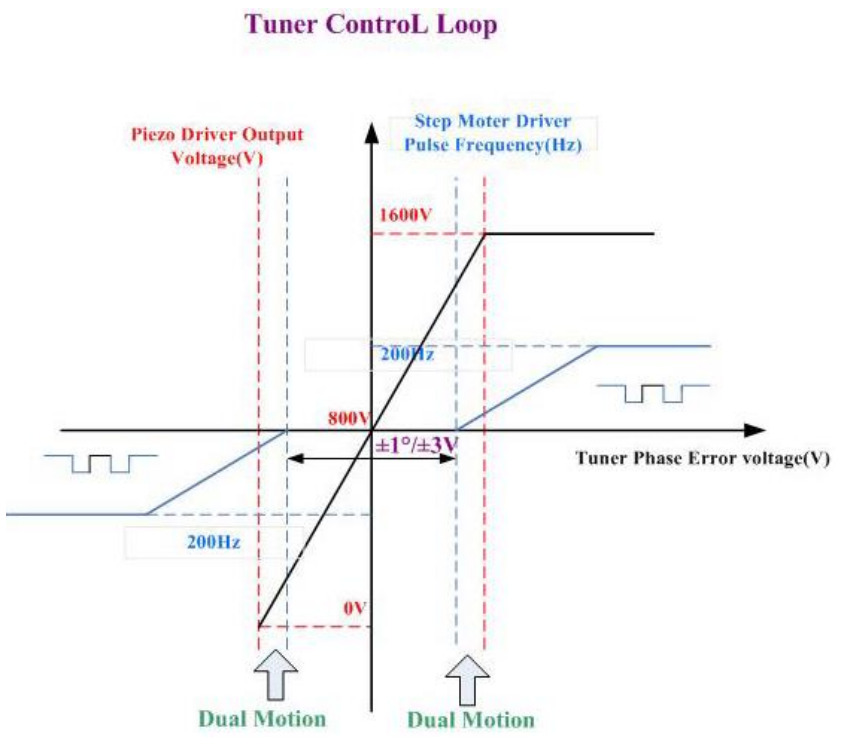

Figure 4: The tuner phase error voltage vs. the stepping motor clock/piezo driver signal.

Figure 4 divides the tuner phase error into four ranges: (1) the stepping motor dead-band and piezo-tuner range, (2) the overlap range, (3) the stepping-motor range and piezo-tuner dead-band, and (4) the tuner-speed-limiting range. Setting like this can reduce motor wear and improve the frequency precision.[4]

When the tuner phase error is within ±1°, the piezo-tuner driver generates a voltage, proportional to the tuner phase error, to change the cavity frequency, while the stepping motor stays in a stop state. When the tuner phase error is in an overlap range, both the piezo-tuner and stepping-motor drivers generate drive signals to change the cavity frequency. When the tuner phase error is in the stepping motor range, only the stepping motor driver clock signal is generated to change the cavity frequency; the driver clock frequency is linearly proportional to the tuner phase error. When the tuner phase error exceeds a limiting setting, the driver clock frequency is limited to a constant, so the moving speed of the stepping motor tuner is also limited. Details of the tuning loop are presented in Figure 5.

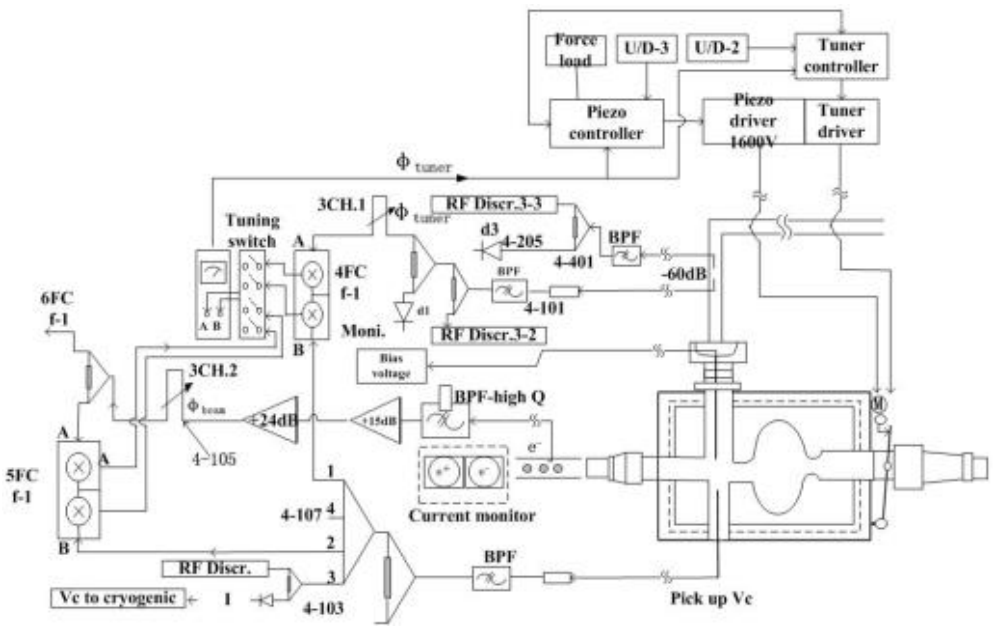

Figure 5: The tuning loop of the KEKB-type SRF cavity of BEPC II.

## 4. The actual working state analysis of tuner system of BEPC II

The main factors that influence the frequency of BEPC II cavity are beam loading, helium pressure and microphonics. The frequency changed by beam loading can be computed by Formula (2).

$$\Delta f = (I_b * \sin\Phi_s / 2V_{rf}) * (R/Q) * f_{rf} \quad (2)$$

R/Q=95.3, $I_b$ is the beam current intensity, $f_{rf}$ is the working frequency of the cavity 499.8 MHz.[5]

The frequency changed by helium pressure is about 250 kHz/bar. The frequency of the cavity changes by about 40Hz due to the external mechanical perturbations and the change of vacuum degree.

According to Formula (2) we know that 900 mA electron beam leads to the cavity detuning by about 14 kHz(@$E_0$=1.89GeV). In order to keep resonance, the tuner force reduces, then the cavity rebounds, the frequency of the cavity decreases to offset the beam impaction. The spring-back displacement of the cavity is about 42.4um. Considering the injection time electron is 10 minutes, the cavity will spring back 71nm/s, the frequency decrease by about 23.4Hz/s. According to Formula (1), 1mm displacement of the motor tuner needs about 9375 pluses. So 1mm displacement of the cavity needs 18750 pulses. Given the injection rate, the tuner needs 1.33 pluses/s.

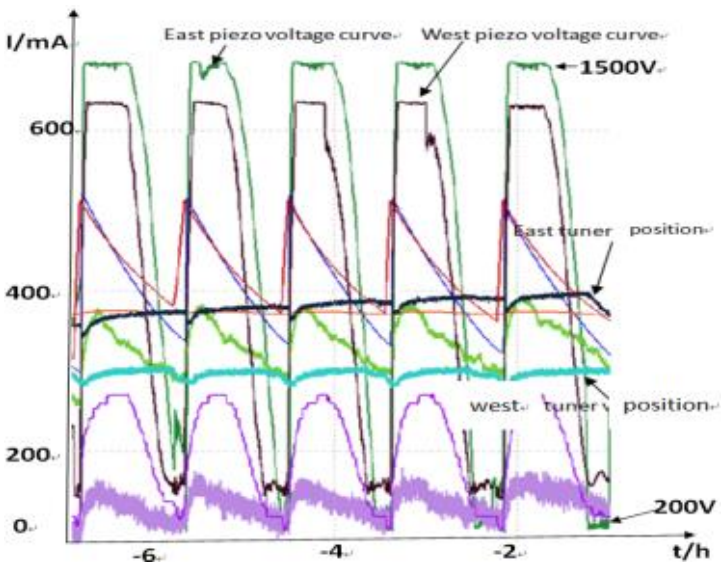

Figure 6: The tuner working state of BEPC II.

As Figure 6 shows, during the process of positron and electron injection, due to the fast injection rate, the phase error is beyond the range of ±1°, only the motor tuner works, the piezo-tuner is in stop state. After the injection is finished, the piezoelectric oscillator voltage goes up to the maximum value quickly. During the start stage, the beam declines, the piezo-tuner and motor tuner both work, but during the last stage, only the piezo-tuner works. To ensure the stability of beams usually set a negative detuning angle.

## 5. The problems of BEPC II tuning system encountered

The reflection power of BEPC II west cavity is abnormal. The minimum reflection power is larger than 10 kW, and the reflection power curve is nearly flat as shown in Figure 7. In order to eliminate the hidden dangers a lot of experiments have been done. At last, it was found that the tuning working point deviated from the resonance point. Adjusting tuning point to the resonance point by moving the phase shifter, the reflection power can be lower than 10 kW. When the cavity voltage is decreased, the reflection power can even decrease to about 2 kW.

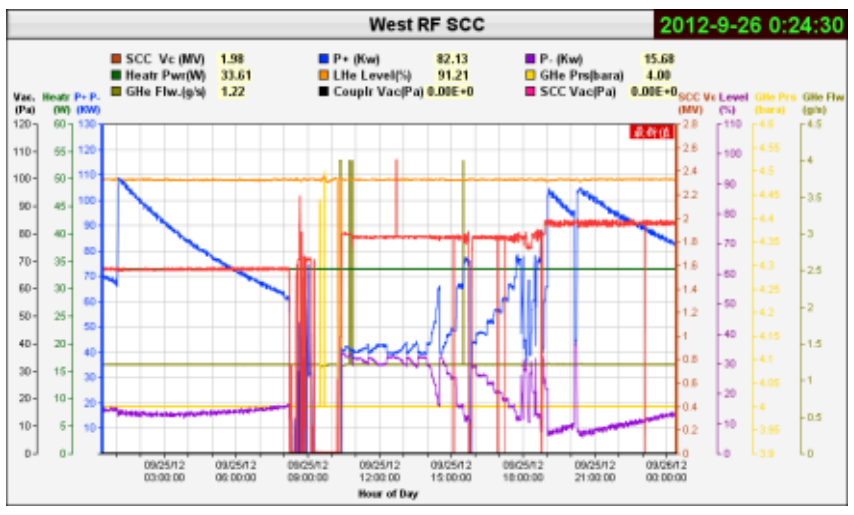

Figure 7: The BEPC II west cavity matching point curve

On March 27, 2012, the electron beam lost frequently, and the motor tuner working position needed constant manual adjustment before beam injection. Figure 8 (a) shows the beam and power curves when the beam is lost. Due to the tuner problem the incident power increased rapidly, the cavity voltage decreased and the beam was lost. Figure 8 (b) shows that when the e-beam decreased, the motor tuner run abnormally and that the tuner position changed suddenly. Through inspection, it was found that the spring screw blocks the motion of tuner arm. When the tuner's push force was greater than the block force, the position of motor tuner changed suddenly.

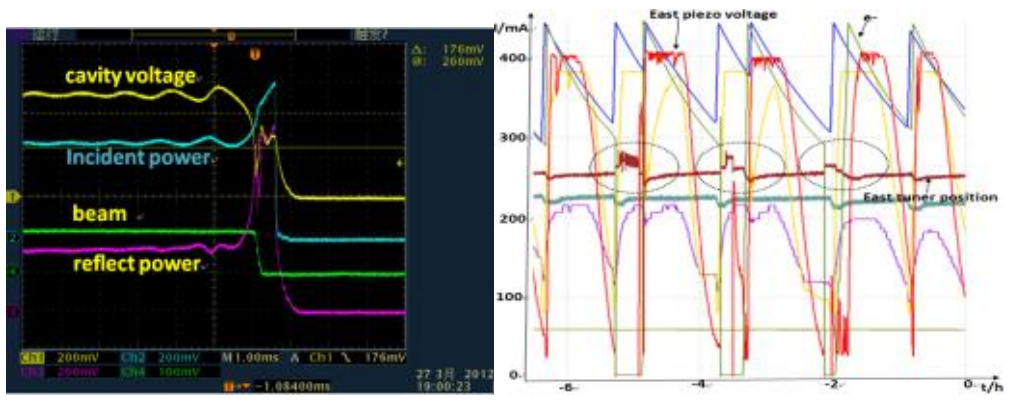

(a) (b)

Figure8: (a) The beam and power curves when the beam was lost; (b) The curves of motor and piezo.

On September 21, 2012，when BEPC II west cavity was run in synchronization mode, the tuning range of tuner was limited. Through inspection, it was found that the pedestal tilted to the south side of the cavity, result in the up part of the tuner arm inclined to the cavity. Through removing the mechanical spacing shim, the tuner could work temporarily. Due to the tuner arms tilt, the cavity might be deformed in the axis. This is very dangerous to the cavity. In such a case, the cavity needs to be warmed up in order to adjust the tuner arm.

## 6. Conclusion

The tuning system of KEKB-type used by BEPC II has worked well during six years on the whole. In order to control the frequency of the cavity better, we need to understand the beam and tuning system. The piezo-tuner needs further research to compensate the frequency of the cavity fast. We have manufactured two sets of tuner and will test them on BEPC II later.

## 7. Acknowledgment

The authors wish to thank the members of IHEP RF group for their perfect cooperation during our work. Special thanks go to KEK experts S. Mitsunobu, T.Furuya, K. Akai for their constant help during these years to us.

## References:

[1] C. Zhang for the BEPCII Team. AN OVERVIEW OF THE BEPCII PROJECT, proceedings of 40th ICFA ABDW 2008.

[2]MI Zheng-Hui etal. Horizontal test for BEPCII 500MHz spare cavity, CPC, 2012, 36(10):996-999

[3] Y. Yamamoto etal.HORIZONTAL TEST FOR CRAB CAVITIES IN KEKB, SRF 2007, WEP27.

[4] M.S.Yeh etal, LOW-LEVEL RF CONTROL SYSTEM FOR THE TAIWAN PHOTON SOURCE, IPAC2011

[5] H. Padamsee et al. RF Superconductivity for Accelerator, second edition, 2007. pp 425-458